\def\d{\mbox{d}}
\documentstyle[pra,aps,amssymb]{revtex}
\begin{document}

\draft
\wideabs{

\title{Sum rule for the pair correlation function}

\author{A. Yu. Cherny\footnotemark[1]}

\address{Bogoliubov Laboratory of Theoretical Physics,
Joint Institute for Nuclear Research, 141980, Dubna, Moscow region,
Russia}

\date{September 17, 2002}

\maketitle

\begin{abstract}
A sum rule has been derived for the static pair correlation function. This rule is the extension of the well-known
equation that relates density fluctuation to compressibility. The obtained sum rule is applied to the Bose and Fermi
ideal gases as well as BCS and Bogoliubov's models.
\end{abstract}
}

\footnotetext[1]{E-mail: cherny@thsun1.jinr.ru}


The well-known sum rule that relates density fluctuation to compressibility of the equilibrium thermodynamic system
plays important role, since it is applicable to description of classical as well as quantum liquids at all densities.
The relation, obtained by Ornstein and Zernike as early as in twenties for a homogeneous system (see, e.g.,
Ref.~\cite{gold} and references therein), reads
\begin{equation}
\int\d^3r\,n[g(r)-1]=\frac{T}{n}\left(\frac{\partial n}{\partial\mu}\right)_T-1,
\label{oc}
\end{equation}
where $g(r)$ is the pair distribution function, which describes the density correlations, and $T$, $\mu$, and $n=N/V$
denote temperature, chemical potential, and density of particles, respectively. The first term in r.h.s. of
Eq.~(\ref{oc}) is proportional to the fluctuations of the number of particles in the Gibbs grand ensemble and can be
connected also with the thermal compressibility: $T(\partial n/\partial\mu)_T/n =T(\partial n/\partial p)_T
=\langle\delta \hat{N}^{2}\rangle/N$, here the first equation is the thermodynamic one, and the second one results from
Eq.~(\ref{Aid}) for $\hat{A}=\hat{N}$ (see below). Below we shall prove that the following generalization is valid for
a thermodynamic system in the equilibrium state:
\begin{eqnarray}
&&\int \d2\,\big[\langle{\hat\psi}^{\dag}(1){\hat\psi}^{\dag}(2){\hat\psi}(2){\hat\psi}(1')\rangle^{\rm as}\nonumber\\
&&\phantom{\int \d2\,\big[}-\langle{\hat\psi}^{\dag}(2){\hat\psi}(2)\rangle^{\rm as}
\langle{\hat\psi}^{\dag}(1){\hat\psi}(1')\rangle^{\rm as}\big]\nonumber\\
&&=T\frac{\partial}{\partial \mu}\langle\hat{\psi}^{\dag}(1)\hat{\psi}(1')\rangle^{\rm as}
-\langle\hat{\psi}^{\dag}(1)\hat{\psi}(1')\rangle^{\rm as},
\label{rule}
\end{eqnarray}
where $\langle\cdots\rangle^{\rm as}$ means the average over the Gibbs canonical grand ensemble in the thermodynamic
limit\cite{note1}, $1=({\bf r}_1,\sigma_1)$ stands for the coordinate and spin indices of a particle, $\int\d2\cdots =
\sum_{\sigma_2}\int\d^{3}r_{2}\cdots$, $\hat{\psi}^{\dag}$ and $\hat{\psi}$ denote the Bose or Fermi field operators.
This relation is correct even for a non-homogeneous system. In the homogeneous case the local density of particles
$\langle{\hat\psi}^{\dag}(2) {\hat\psi}(2)\rangle^{\rm as}$ becomes constant, and r.h.s. of Eq.~(\ref{rule}) depends
only on ${\bf r}_1-{\bf r}'_1$, which allows us to employ the Fourier transformation when investigating a specific
model. We stress that the sum rule contains {\it asymptotic limiting values} of the correlation functions, and, hence,
one can replace the condensate operators by $c$-numbers in the case of Bose system below the temperature of the
Bose-Einstein condensation, as was {\it proved} by Bogoliubov~\cite{bogquasi}. The relation~(\ref{rule}) is the
generalization of Eq.~(\ref{oc}), which is obtained from Eq.~(\ref{rule}) by setting $1=1'$ and using the formula
$\sum_{\sigma_1} \langle\hat{\psi}^{\dag}(1) \hat{\psi}(1)\rangle=n$; here we put by definition $g(|{\bf r}_2-{\bf
r}_1|)=\sum_{\sigma_1,\sigma_2}\langle{\hat\psi}^{\dag}(1) {\hat\psi}^{\dag}(2) {\hat\psi}(2) {\hat\psi}(1)
\rangle^{\rm as}/n^2$ for the pair distribution function.

The proof is rather simple. Let us consider a subsystem of volume $V$ embedded in a system of volume $V_0$ with which
the subsystem can exchange energy and particles. It follows from the definition of the number operator
$\hat{N}=\int_{V}\d2\,\hat{\psi}^{\dag}(2) \hat{\psi}(2)$ of the subsystem and the commutation relation
$[\hat{N},\hat{\psi}(1')]=-\hat{\psi}(1')$
\begin{eqnarray}
&&\int_{V}\d2\,\big[\langle{\hat\psi}^{\dag}(1){\hat\psi}^{\dag}(2){\hat\psi}(2){\hat\psi}(1')\rangle \nonumber\\
&&\phantom{\int \d2\,\big[}-\langle{\hat\psi}^{\dag}(2){\hat\psi}(2)\rangle
\langle{\hat\psi}^{\dag}(1){\hat\psi}(1')\rangle\big]\nonumber\\
&&=\langle\hat{\psi}^{\dag}(1)\hat{\psi}(1')(\hat{N}-N)\rangle - \langle\hat{\psi}^{\dag}(1)\hat{\psi}(1')\rangle,
\label{auxil}
\end{eqnarray}
here the coordinates ${\bf r}_1$ and ${\bf r}_1'$ in the indices $1$ and $1'$ range over the subsystem only. Let us
perform the thermodynamic limit $V_0\to\infty$, $n={\rm const}$; then the subsystem can be considered as the grand
ensemble. Taking after that the limit $V\to\infty$ and using the general relation for the grand ensemble
($[\hat{A},\hat{N}]=0$ is assumed)
\begin{equation}\label{Aid}
\langle\hat{A}\hat{N}\rangle-\langle\hat{A}\rangle\langle\hat{N}\rangle
=T{\partial\langle\hat{A}\rangle}/{\partial\mu}
\end{equation}
for $\hat{A}=\hat{\psi}^{\dag}(1)\hat{\psi}(1')$, we arrive at Eq.~(\ref{rule}). Note that performing the limit
$V_0\to\infty$ is the essential point of the proof, for the two-body correlation function in Eq.~(\ref{auxil}) can
contain, apart from its asymptotic value, terms of the order of $1/V_0$, which do not tend to zero at large distances
between ${\bf r}_1,\ {\bf r}_1'$ and ${\bf r}_2$ and could give a finite contribution to the integral in
Eq.~(\ref{auxil}). The identity (\ref{Aid}) can be verified directly from the expression of averages
$\langle\hat{A}\rangle={\rm Tr}\,(\hat{A}\hat{\rho})$ in the canonical grand ensemble with the density matrix
$\hat{\rho}=\frac{1}{Z}\exp[-(\hat{H}-\mu\hat{N})/T]$, here $Z={\rm Tr}\,\exp[-(\hat{H}-\mu\hat{N})/T]$ is the grand
partition function~\cite{note}.

It is not difficult to convince ourselves that the sum rule (\ref{rule}) is satisfied for the ideal Fermi gas or Bose
one above the critical temperature. Indeed, using the Wick's theorem and performing the Fourier transformation yield
$\mp n_{{\bf k}\sigma}^{2}=T\partial n_{{\bf k}\sigma}/\partial \mu - n_{{\bf k}\sigma}$, which is obviously fulfilled
for the occupation numbers. Below the critical temperature divergences appear in Eqs.~(\ref{oc}), (\ref{rule}) and
(\ref{Aid}). This is due to the fact that the chemical potential of the ideal Bose gas becomes zero and independent of
the density and temperature in the thermodynamic limit; as a consequence, the derivatives with respect to the chemical
potential go to infinity. In particular, that means that the fluctuations of the total number of particles are
non-thermodynamic in the grand ensemble: $\langle\delta \hat{N}^{2}\rangle/N \to \infty$ in the limit $V\to\infty$
instead of the normal behaviour $\langle\delta \hat{N}^{2}\rangle\propto N$, which Eq.~(\ref{Aid}) implies for
$\hat{A}=\hat{N}$; this is well-known anomaly of the ideal Bose gas~\cite{box-id}.

The correlation functions in the exactly solvable model of Bardeen, Cooper and Schrieffer can be evaluated also with
the Wick's theorem~\cite{bloch}, as the Hamiltonian can be approximated (exactly in the thermodynamic limit) by the
quadratic form of the Fermi operators. It is not difficult to see that in this case the sum rules~(\ref{oc}) and
(\ref{rule}) are violated. The violation of Eq.~(\ref{oc}) was shown by J. Bell, who revealed parallels between this
problem and the well-known problem of the gauge-invariant Meissner effect in BCS model~\cite{bell}.

Let us now examine the Bogoliubov model~\cite{bog47} for spinless bosons. At first, we select the $c$-number parts in
the field operators $\hat{\psi}^{\dag}=\sqrt{n_0}+\hat{\vartheta}^{\dag}$ and $\hat{\psi} =\sqrt{n_0}
+\hat{\vartheta}$. Then the correlation functions are calculated easily with the Wick's theorem for the operators
$\hat{\vartheta}^{\dag}$ and $\hat{\vartheta}$, since the model Hamiltonian is the quadratic form of the Bose
operators~\cite{bog47}. Thus, in addition to obvious relations $\langle\hat{\vartheta}\rangle
=\langle\hat{\vartheta}^{\dag}\rangle=0$, we have for the triple averages $\langle\hat{\vartheta}^{\dag}
\hat{\vartheta} \hat{\vartheta}\rangle= \langle\hat{\vartheta}^{\dag}\hat{\vartheta} ^{\dag}\hat{\vartheta}\rangle=0$.
Performing the Fourier transformation of Eq.~(\ref{rule}) yields for $k=0$
\begin{equation}
2n_{0}\lim_{k\to 0}(\psi_{\bf k}+n_{\bf k})=-n_{0}+T\frac{1}{(\partial \mu/\partial n_{0})},
\label{k0}
\end{equation}
and for $k\not=0$
\begin{equation}
|\psi_{\bf k}|^{2}+n_{\bf k}^{2}=-n_{\bf k}+T\frac{\partial n_{\bf k}}{\partial\mu}.
\label{knot0}
\end{equation}
Here $\psi_{\bf k}=\langle\hat{a}_{\bf k}\hat{a}_{-{\bf k}}\rangle$ and $n_{\bf k}=\langle\hat{a}^{\dag}_{\bf
k}\hat{a}_{\bf k}\rangle$ stand for the normal and anomalous averages. The relation~(\ref{k0}), which corresponds to
the condensate density, is satisfied exactly; this can be verified with the expressions of paper~\cite{bog47} with
$\mu=n_0\tilde{\Phi}(0)$ (here $\tilde{\Phi}(0)$ is the zero Fourier component of the interaction potential). At the
same time, one can readily see that Eq.~(\ref{knot0}) is not fulfilled at zero temperature due to the terms $|\psi_{\bf
k}|^{2}$ and $n_{\bf k}^{2}$, resulting from the four-boson average $\langle\hat{\vartheta}^{\dag}
\hat{\vartheta}^{\dag} \hat{\vartheta} \hat{\vartheta}\rangle$. At non-zero temperatures we arrive at obvious
divergence in the l.h.s. of Eqs.~(\ref{oc}) and (\ref{rule}) due to those terms, since $|\psi_{\bf k}|^{2}\sim n_{\bf
k}^{2}\sim 1/k^4$ at $k\to0$ according to the Bogoliubov's ``$1/k^2$" theorem~\cite{bogquasi}. Note that the three- and
four-boson averages are supposed to be of the next order of smallness in comparison with $|\psi_{\bf k}|$ and $n_{\bf
k}$ within the Bogoliubov theory; in particular, they are neglected in the Bogoliubov's Hamiltonian~\cite{bog47}. From
Eq.~(\ref{knot0}) one can conclude that evaluation of the four-boson averages is not consistent in the framework of the
Bogoliubov model and, consequently, exceeds its accuracy. It is worth noting that in calculating the pair distribution
function $g(r)$ Bogoliubov restricted himself by the two-boson terms $\psi_{\bf k}$ and $n_{\bf k}$ only~\cite{boglec}.

In conclusion, the generalization~(\ref{rule}) of the well-known relation~(\ref{oc}) is proposed. Violation of the sum
rule~(\ref{rule}) may indicate an anomaly in a model or point out limits of its validity. In particular, we show that
calculation of the four-bosons averages is beyond the accuracy of the Bogoliubov model.

This work was supported by the RFBR grant 01-02-17650. Discussions with V.B.~Priezzhev are gratefully acknowledged.


\end{document}